# On the microscopic mechanism of ion-extraction of a gridded ion propulsion thruster

D.Kirmse

Institute for Aerospace Engineering, Technical University Dresden, Germany

**Abstract:** The following paper includes a physical microscopic particle-description of the phenomena and mechanisms that lead to the extraction of ions with the aim to generate thrust. This theoretical treatise arose from the intention to visualize the behavior of the involved particles under effect of the involved electrical fields. By this way, an underlying basis for experimental investigations of the work of an ion thruster should be formed. So a foundation was created, which explains the ion extracting and so thrust generating function of an ion thruster. The theoretical work was related to the Radio-frequency Ion Thruster (RIT). But the model worked out can be generalized for all thruster types that use electrostatic fields to extract positively charged ions.

The framework for these reflections is a thesis [1], which includes conception, construction and characterization of the micro-newton ion thrusters RIT-2 and RIT-4. At the attempt to find a theoretical basis for the function behavior of such an ion thruster one fact became clear; there is no sufficient treatment of the thruster behavior with respect to a cause for ion extraction. A lot of experimental work was done with Radio-frequency Ion Thrusters, this includes also a mathematical linkage of the observed values [2], [3]. But there is no conclusive and holistic physical picture of the particle behavior that leads to the macroscopic behavior of extracting ions out of the thruster.

The following description is based only on first principles; on the electric field, the electric charge as both source and target of electric field, and on the thermal movement of particles. The primary driving force is the resulting electric field between regions with different density of charges. Without any falsification it can be considered a time-hierarchic structure of the events inside the thruster. Events, which proceed simultaneously in real, were separately treated for a better understanding of the single processes.

Initially, the key effect is the electrical interaction between plasma grid and charged particles of plasma inside the thruster. All gridded ion thrusters have the same principle of function to create thrust. An electric or electromagnetic field causes the production of plasma inside the thruster by gas discharge. In case of a RIT an oscillating current inside a coil generates a cyclic closed eddy-field in a discharge vessel, which is surrounded by the coil. This alternating field accelerates free electrons. The electrons impact in neutral gas atoms and ionize the atoms via transfer of inner energy; plasma is produced (fig.1). The plasma grid is in direct contact to the plasma (fig.1). This plasma grid is the source of an electro-static field, that extracts and accelerates positively charged ions to generate thrust; so far the experimental observation. The microscopic reason for this behavior will be explained in the following by a step by step treatment of the involved particles under influence of electric fields, plus the effect of thermal movement.

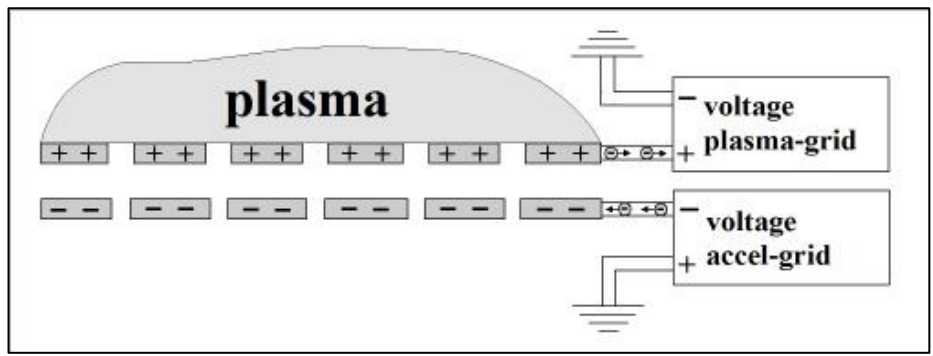

1 Scheme of the configuration of plasma and acceleration grid including the voltage supply

The basic condition of all these mechanisms and effects is the presence of charged particles. This is realized by electron impact ionization and the resulting generation of plasma. In figure (2a) the production of electrons and positively charged ions is shown. Now the plasma grid is able to influence the plasma. The plasma grid is connected to the positive end of a voltage-source. A voltage-source separates the two species of charges. The underlying mechanism is the metallic conductivity with only electrons as moveable charged particles. So inside the voltage-source electrons are displaced with the result of separation in a region of electron-depletion (positive end) and a region of electron-accumulation (negative end). The electron-depletion leads to an effective positive charge, which is source for an electron-attracting electric field. Conversely, effective negative charge of electron-accumulation is source for an electron-repulsive electric field. If a conductor is connected to one of the ends of a voltage-source, this resulting electric fields cause the displacement of electrons in the connected conductor, too. In this particular case the plasma grid is connected to the positive end, so its resulting electric field attracts the moveable electrons of the grid. The consequence is an electron-current from the grid into the positive end of the source and a resulting depletion of electrons also in the grid. This current is existent until the effective (positive) charge-density in the grid is equal to the density in the positive end. If the densities are equal, there is no effective electric field between grid and positive end; the electron-current disappears. The result of this process is a plasma grid with a positive surplus charge-density; now the grid is a source for electron-attracting electric field, especially towards the plasma. Inside the plasma; the free electrons are attracted and the positive ions are repulsed by this field (fig. 2b). The plasma electrons reach the grid and they are conducted into the positive end of the voltage-source. This process continues until the effective density of electrons inside plasma and grid is equal. And without different charge-densities between plasma and grid there is no reason for a resulting electric field that influences the charged particles in the plasma (fig. 2c).

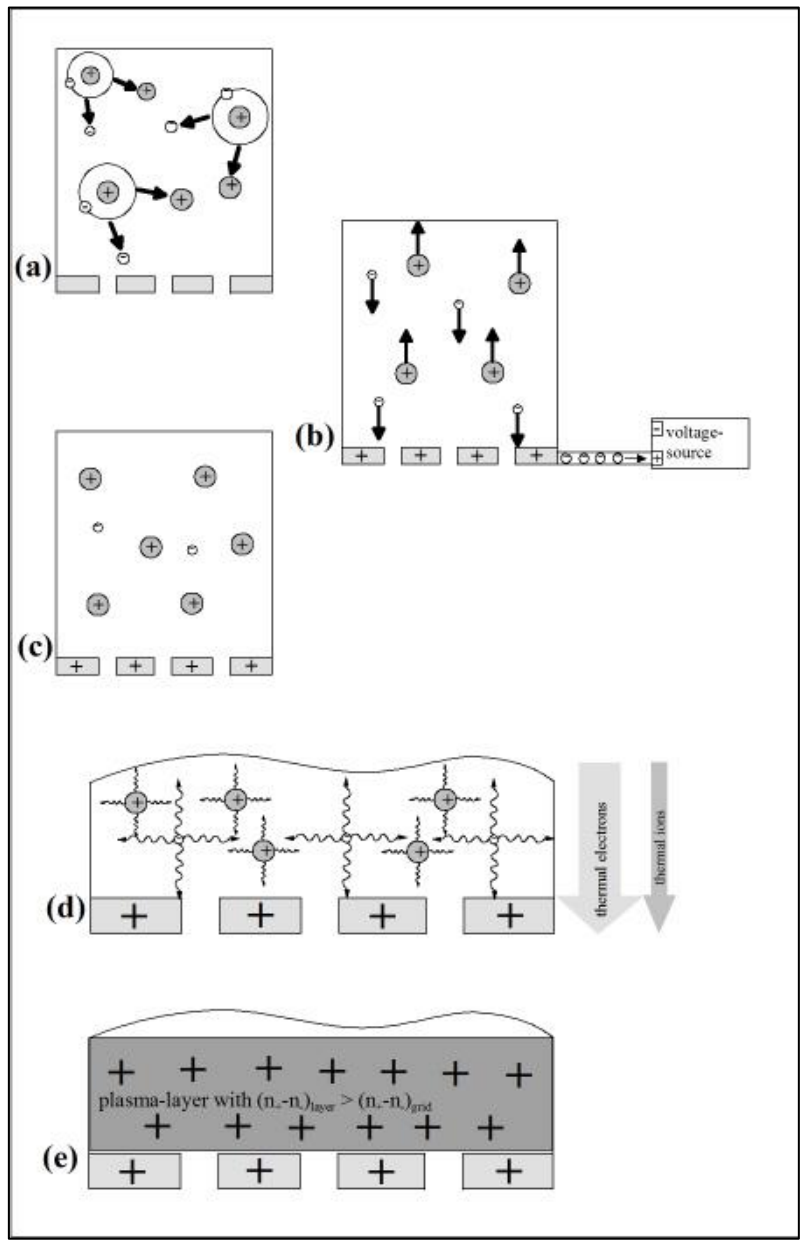

2 Mechanisms that lead to ion extraction.

At this point an electrostatic equilibrium is achieved. The charge-densities are in balance and no resulting electric fields appear; the plasma particles remain un-influenced. Without any additional effects but the electro-static there is no explanation or reason for ion extraction out of the plasma.

But there is one more physical process that has to be considered: the thermal movement of the charged plasma particles. Without falsification one can assume all particles in a thermal equilibrium. Hence, the particles have equivalent kinetic energies under this condition. However, the mass of the electrons is much smaller than the mass of the ions. This implies that the thermal velocity of the electrons is much higher than the velocity of the ions; in the same ratio the ions are heavier than the electrons. Hence, the current of particles (which is proportional to the velocity of particles) is for electrons higher than for ions in every direction of space. And especially, the current directed towards the grid is for electrons higher than for ions (fig. 2d). The grid again conducts also these thermal electrons into the positive end of the voltage-source. With the result: the thermal movement causes an additional depletion of electrons in a plasma-layer close to the plasma grid (fig. 2e). And this difference between effective charge-densities of plasma-layer and plasma grid is the cause for a resulting electric field between layer and grid, which is repulsive to ions and forces ions from inside the plasma-layer towards and through the plasma grid (fig. 3), and which is attractive to electrons and counteracts the thermal electron-current toward the grid.

The result is a particle-based mechanism, combined of electro-static interaction and thermal movement that explains the reason of ion-extraction out of an electro-static ion-thruster. The actual mechanism of ion-extraction is a closed-cycle self-balancing equilibrium: difference in charge-density causes the extraction of ions and the counteracting of thermal electron current, ion extraction reduces difference in charge-density, this enables the current of thermal electrons from plasma-layer to grid and the difference in charge-density is renewed; and the cycle of action starts from the beginning.

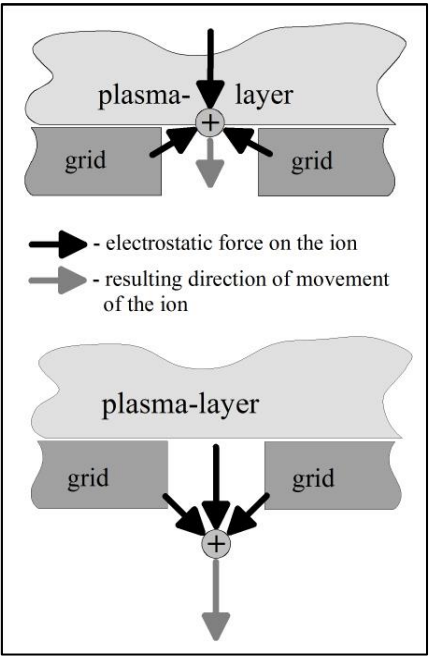

3 Single electrostatic forces from plasma layer and plasma grid and resulting electrostatic force on the ion.

The difference of charge density in the plasma layer and the plasma grid and the resulting electric field between them of course also leads to a difference of energy for ions that remove from the layer or the grid. This difference of energy is commonly called plasma potential. The value of the plasma potential is usually some eV above the electric potential of the plasma grid; depending on the quantity of additional thermal depletion of electrons in the plasma layer.

This correlation between electrons that are lead away by the plasma grid and the extracted ions from the plasma is interesting for experimental investigation, too. The current of electrons can be simply measured inside the voltage source. And because the electron current is identic to the current of extracted ions, there is an opportunity to indirectly investigate the number of ions that leave the thruster.

As outlined, the reason for ion extraction is an electric field starting from a plasma layer. Causal for this behavior is the charged plasma grid and the thermal movement; so the plasma grid is responsible for ion extraction. The acceleration (accel) grid, placed in front of the plasma grid towards the exit of the thruster has no effect on ion extraction. Because the field effect from the accel grid on the ions passing through the accel grid is equal but opposite before and behind the grid, the force on passing ions is nullified (fig. 4). So the accel grid has no extracting (and accelerating) effect. Its only effect is to focus the extracted ion beam.

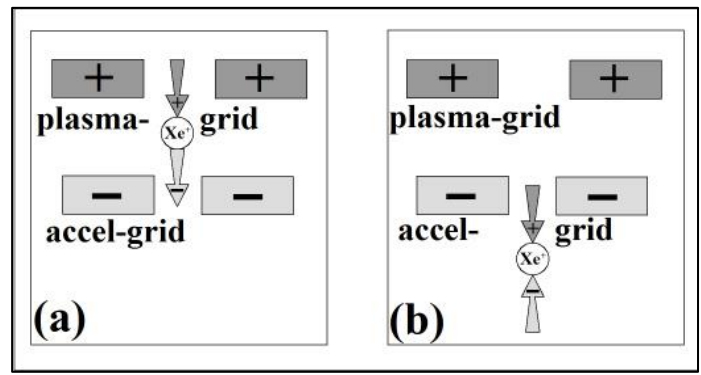

**4 Electrostatic force from plasma grid and accel grid on a positive ion passing through the grids.**

---

Mathematical relations:

Difference in charge density as source of an electrostatic field

$$E \sim (n_+ - n_-) \cdot q \ ; for\ n_+ > n_-$$

Conditions without thermal movement

$$(n_+ - n_-)_{grid} = (n_+ - n_-)_{plasma}$$

Conditions with thermal movement including appearance of plasma potential

$$n_{ion,layer} - n_{electron,layer} = n_{diff,layer} > n_{diff,grid} = n_{ion,grid} - n_{electron,grid}$$

$$v_{therm,electron} - v_{therm,ion} \sim n_{diff,layer} - n_{diff,grid} \sim V_{plasma}$$

$$E - electrostatis\ field, n - charge\ density, v - velocity, V_{plasma} - plasma\ potential$$

---